\documentclass{article}
\usepackage{spconf,amsmath,graphicx}
\usepackage{soul}
\usepackage[colorlinks=true, linkcolor=blue, urlcolor=blue, citecolor=blue, filecolor=blue]{hyperref}

\title{An LSTM Feature Imitation Network \\ for Hand Movement Recognition from sEMG Signals}
\name{Chuheng Wu$^1$, S. Farokh Atashzar$^1$, Mohammad M. Ghassemi$^2$, Tuka Alhanai$^3$*\thanks{*T.A. is corresponding author (tuka.alhanai@nyu.edu). Support provided by Center for AI and Robotics (CAIR) at New York University Abu Dhabi (funded by Tamkeen under NYUAD Research Institute Award CG010), and NSF Award \#2320050. Technical support provided by High-Performance Computing team at NYUAD and NYU. Code available at: \href{https://github.com/x-labs-xyz/aaai-25-fin-lstm-emg}{https://github.com/x-labs-xyz/aaai-25-fin-lstm-emg}}}
\address{\textit{$^1$New York University, USA;} \textit{$^2$Michigan State University, USA}; \\  
\textit{$^3$Center for AI \& Robotics, New York University Abu Dhabi, UAE}} 
\begin{document}
%
\maketitle
\begin{abstract}

Surface Electromyography (sEMG) is a non-invasive signal that is used in the recognition of hand movement patterns, the diagnosis of diseases, and the robust control of prostheses. Despite the remarkable success of recent end-to-end Deep Learning approaches, they are still limited by the need for large amounts of labeled data. To alleviate the requirement for big data, we propose utilizing a feature-imitating network (FIN) for closed-form temporal feature learning over a 300ms signal window on Ninapro DB2, and applying it to the task of 17 hand movement recognition. We implement a lightweight LSTM-FIN network to imitate four standard temporal features (entropy, root mean square, variance, simple square integral). We observed that the LSTM-FIN network can achieve up to 99\% R2 accuracy in feature reconstruction and 80\% accuracy in hand movement recognition. Our results also showed that the model can be robustly applied for both within- and cross-subject movement recognition, as well as simulated low-latency environments. Overall, our work demonstrates the potential of the FIN modeling paradigm in data-scarce scenarios for sEMG signal processing.
\end{abstract}
%
%
\section{Introduction}
\label{sec:intro}
Surface Electromyography (sEMG) is a valuable, non-invasive indicator of electro-physiological activity widely used in human-computer interfaces and prosthetic control \cite{merletti1989new}. However, interpreting sEMG signals is challenging due to their non-linear nature \cite{farry1996myoelectric} and case dependent application \cite{li2019research}.

Deep learning methods have demonstrated high-accuracy sEMG signal recognition \cite{liu2022novel}. However, end-to-end models are highly data-dependent \cite{waris2018effect}, requiring extensive data to adapt to diverse scenarios \cite{hargrove2006effect}. Feature Engineering offers an alternative approach, mathematically mapping high-dimensional signal windows to relevant features representations \cite{phinyomark2018emg}. Common sEMG features include time domain features \cite{waris2018effect}, frequency domain features \cite{zhuang2020comprehensive}, time-frequency domain features \cite{waris2020multiday}. Researchers have incorporated these principles into deep learning approaches; CNNs excel at extracting spatial features from sEMG spectrograms, while LSTM networks effectively exploit temporal features \cite{bao2020cnn}. However, the learned feature representations are abstract and typically beyond a closed-form mathematical description (i.e. uninterpretable) \cite{park1998emg, srhoj2022feature, li2022interpretable}. In this work, we aim to demonstrate the viability of pre-training networks with an explicit set of manually selected time-domain features, taking advantage of knowledge distilled from feature engineering efforts while also balancing the flexibility provided by deep learning methods to tune and adapt learned representations.

A common solution to data scarcity is to replace the end-to-end training paradigm with feature-to-end transfer learning \cite{zhuang2020comprehensive}. 
While the feature-to-end transfer learning approach has been actively explored, the end-to-feature modeling paradigm has been less explored, but is increasingly finding appeal in the context of self-supervised learning \cite{mohamed2022self}. A novel approach in the domain of feature-to-end transfer learning are Feature Imitation Networks (FIN), which have shown state-of-the-art performance on several biomedical signal and image processing tasks \cite{saba2022feature, min2023feature}. However, there has been no prior work to evaluate the potential of FINs for sEMG signal processing tasks. In this paper, our aim is to explore the end-to-feature learning approach by utilizing a set of established time-domain sEMG features (we emphasize that this is distinct from the conventional end-to-end or feature-to-end learning approaches). Specifically, our aim is to apply a temporal network architecture (LSTM), in the form of a FIN, to imitate time-domain features \cite{saba2022feature, min2023feature}. The pre-trained FINs are then fine-tuned for downstream classification tasks (by feeding into a feature-to-end model).

%




\section{Data and Methodology}
\label{sec:Data}
\subsection{Data Acquisition and Processing}
We utilized the NinaPro dataset \cite{atzori2014electromyography}, Exercise B of the DB2 database, which includes 17 hand movements (8 isometric and isotonic hand configurations, 9 basic wrist movements). The dataset comprises sEMG data from 12 electrodes on the right forearm of 40 subjects (12 female, 28 male). Each movement was repeated 6 times with a 5-second hold and 3-second rest, sampled at 2 kHz.

The target variable was a one-hot vector representing the 17 hand movements. Following the dataset's split method \cite{atzori2014electromyography}, repetitions 2 and 5 were designated as the test set, while 1, 3, 4, and 6 were for training. We further divided this into ``within-subjects" (1-25) and ``cross-subjects" (26-40) sets to explore transfer learning ability. For subjects 1-25, repetitions 1, 3, and 4 were used for pre-training, and 6 for fine-tuning. For subjects 26-40, repetitions 1, 3, 4, and 6 were used for fine-tuning. The test set comprised repetitions 2 and 5 from all subjects.


Zero-mean and unit-variance normalization was applied to each channel \cite{sun2021temporal}. Input data was generated using a sliding window of 300ms with a 10ms stride, resulting in a 600x12 window size (600 sample points, 12 electrode channels). This window size balances pattern learning and minimizes delays in real-time scenarios \cite{sun2021temporal}.

\subsection{Features} 
We focused on four widely explored, complementary time-domain features: Entropy (ENT), Root Mean Square (RMS), Variance (VAR), and Simple Square Integral (SSI) \cite{waris2018effect}. RMS represents constant force and non-fatiguing contraction \cite{gennisson2005human}; VAR measures signal variation \cite{negi2016feature}; Entropy represents signal irregularity \cite{cao2004detecting}; and SSI represents accumulated energy in a window \cite{du2004temporal}. These features were chosen for their significant impact on classification accuracy and complementary information capture. Z-score normalization was applied to each feature set. 

\begin{figure}[t]
    \centerline{\includegraphics[scale=0.4]{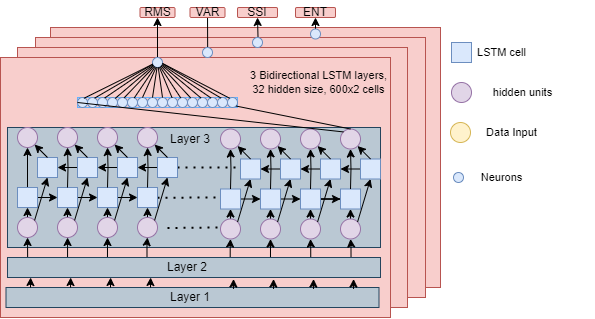}}
    \caption{Bi-directional Feature Imitating LSTM Neural Network; 3-layer structure, each layer has 600 forward-backward connected LSTM cells. Each LSTM-FIN learns to generate one of four features (RMS, VAR, ENT, SSI).}
    \label{fig:LSTM}
\end{figure}

\section{Model Architecture}
\label{sec:model}
Our architecture consists of two parts: an upstream bi-directional LSTM Feature Imitation Network (FIN) that imitates time-domain features, and a downstream classifier that maps these features to hand movements.

\subsection{Feature Imitating LSTM and Augmentation}
We utilize Bi-LSTMs for the FIN architecture due to their effectiveness in extracting temporal features in time series regression tasks \cite{sun2021temporal}, and their ability to learn both forward and backward relationships \cite{ma2021Bi-LSTM}. For each electrode channel, a one-dimensional Bi-LSTM is employed with an input length of 600, aligning with the window size. After experimentation, we adopted a three-layer architecture with a hidden size of 32, balancing accuracy and training efficiency. The LSTM output feeds into a fully connected perceptron to generate a single feature output. The total number of trainable parameters in the LSTM is 22,508. The architecture is shown in Fig. \ref{fig:LSTM}.

\begin{figure}[b]
    \centerline{\includegraphics[width=0.5\textwidth]{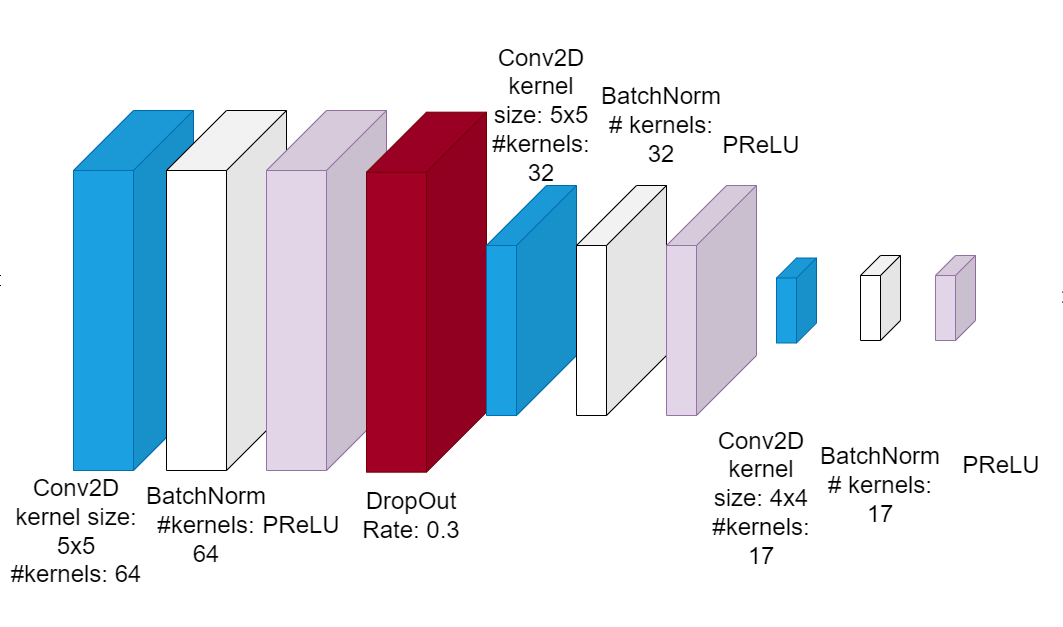}}
    \caption{Downstream neural network architecture for recognition, mapping from augmented features to classifications}
    \label{fig:CNN}
\end{figure}

To improve the classifier's accuracy and robustness, we introduce feature augmentation. We treat the four features (ENT, RMS, VAR, SSI) as distinct channels due to their significant differences in absolute values. For each sampling window, we generate 12 one-dimensional corresponding features for each feature channel. We replicate each feature 11 times and introduce Gaussian noise (mean 0, standard deviation 0.01) \cite{wang2018data}. This results in a $4\times12\times12$ tensor. 

\subsection{Downstream Classifier}
The augmented feature tensor is processed by a three-layer network. Each layer contains a 2D convolutional layer, a batch normalizing layer, and a PReLU activation layer. A dropout layer (rate 0.3) is added between the first and second modules to randomly discard outlier features. The output layer consists of 17 neurons, corresponding to the 17 hand movements (Fig. \ref{fig:CNN}).

\section{Experiments}
\subsection{Exp. $\#$1: Feature Imitating Experiment} 
The LSTM-FIN was trained using an Adam optimizer (initial learning rate 0.001, weight decay $2 \times 10^{-5}$) with early stopping to prevent overfitting. Evaluation metrics included R-squared (R²) and Mean Absolute Percentage (MAP), comparing the model's output with ground truth feature values.

\begin{table}[h]
  \centering
  \caption{LSTM-FIN Model Performance Results}
  \label{tab:FIN_performance}
    \setlength{\tabcolsep}{12pt}
    \begin{tabular}{|c|cc|}
  \hline
    \textbf{Feature} & \textbf{$R^2$} & \textbf{MAP} \\
    \textbf{} & mean (std) & mean (std) \\
    \hline
    ENT & 0.98 (±0.10) & 95.17\% (±11.78\%) \\
    RMS & 0.99 (±0.08) & 95.33\% (±9.47\%) \\
    SSI & 0.96 (±0.14) & 88.62\% (±14.43\%) \\
    VAR & 0.98 (±0.12) & 92.97\% (±13.06\%) \\
    \hline
\end{tabular}
\end{table}

As shown in Table \ref{tab:FIN_performance}, the model demonstrated strong performance, with R² values exceeding 0.96 and MAP values above 88\%. These results highlight the robustness of LSTM-FINs in learning closed-form feature representations.



\subsection{Exp. $\#$2: Downstream CNN Training}
\label{subsec: CNN}
We evaluated the performance of both Feature Engineering and Deep Learning, to be specific: (1) \textbf{SVM}: A traditional feature engineering approach using an SVM model with a Gaussian kernel, trained on ground-truth features; (2) \textbf{CNN}: CNN models trained on ground-truth features, implemented in two variations, (a) \textit{CNN-I} with multi-subject training using repetitions 1, 3, 4, and 6 from subjects 1 to 25, and (b) \textit{CNN-II} with single-subject training using repetitions 1, 3, 4, 6 from one of the 40 subjects. Both testing sets used repetitions 2 and 5 of subjects 1 to 40, with subject-by-subject evaluation.

CNN-II outperforms SVM on all-subjects by over 16\%, indicating the CNN’s learning advantage over traditional feature engineering models (see Table \ref{tab:model_performance}). Moreover, we observe CNN-I achieves low accuracy ($<$ 30\%), implying feature representations are distinct between subjects.

\subsection{Exp. \#3: Downstream Fine-tuning Experiments}
We combined the LSTM-FIN and CNN-II models in four steps:  (1) Pre-train the LSTM-FIN and CNN classifier separately with original data and ground-truth features. (2) Initialize the encoder (LSTM-FIN) and decoder (CNN) with pre-trained weights from corresponding subjects. (3) Integrate encoder and decoder models, enabling weight back-propagation on both parts. (4) Reload the fine-tuning set and fine-tune the model as a whole, adapting to specific subjects for subject-by-subject evaluation. We compare the performance of the LSTM-FIN+CNN-II with a LSTM+CNN-II network (which uses the same architecture and excludes FIN pre-training). We also compare with the SVM model from Section \ref{subsec: CNN}. 


\begin{table}[t!]
  \centering
  \caption{Results of Model Performance}
  \label{tab:model_performance}
  \small
  \setlength{\tabcolsep}{4pt}
  \begin{tabular}{|l|c|c|}
    \hline
    \textbf{Inputs} & \textbf{Model} & \parbox[c]{1.5cm}{\centering \textbf{Accuracy}}\\
    \hline
    \multicolumn{3}{|l|}{\textit{Within-Subject (1-25)}} \\ \hline
    ENT, RMS, SSI, VAR & CNN-I & 29.40\% ± 2.00 \\
    ENT, RMS, SSI, VAR & CNN-II & 64.64\% ± 7.70 \\
    ENT, RMS, SSI, VAR & SVM & 53.59\% ± 6.47 \\
    Raw sEMG & LSTM+CNN-II & 34.57\% ± 8.35 \\
    \textbf{Raw sEMG} & \textbf{LSTM-FIN+CNN-II} & \textbf{65.70\% ± 7.50} \\
    \hline
    \multicolumn{3}{|l|}{\textit{Cross-Subject (26-40)}} \\ \hline
    ENT, RMS, SSI, VAR & CNN-I & 25.13\% ± 1.94 \\
    ENT, RMS, SSI, VAR & CNN-II & 77.90\% ± 7.19 \\
    ENT, RMS, SSI, VAR & SVM & 52.91\% ± 7.70 \\
    Raw sEMG & LSTM+CNN-II & 45.41\% ± 6.33 \\
    \textbf{Raw sEMG} & \textbf{LSTM-FIN+CNN-II} & \textbf{80.06\% ± 6.21} \\
    \hline
    \multicolumn{3}{|l|}{\textit{All-Subject (1-40)}} \\ \hline
    ENT, RMS, SSI, VAR & CNN-I & 27.80\% ± 1.98 \\
    ENT, RMS, SSI, VAR & CNN-II & 69.61\% ± 9.86 \\
    ENT, RMS, SSI, VAR & SVM & 52.94\% ± 8.42 \\
    Raw sEMG & LSTM+CNN-II & 38.64\% ± 9.25 \\
    \textbf{Raw sEMG} & \textbf{LSTM-FIN+CNN-II} & \textbf{71.80\% ± 9.89} \\
    \hline
  \end{tabular}
\end{table}

Notably, CNN-II, LSTM+CNN-II, and LSTM-FIN+CNN-II models perform significantly better in cross-subject recognition compared to within-subject recognition (77.90\% vs. 64.64\%, 45.41\% vs. 34.57\%, and 80.06\% vs. 65.70\%, respectively). This improvement may be attributed to the inclusion of more repetitions in cross-subject experiments, providing more diverse data and improving the model's ability to generalize across inter-repetition samples. These results demonstrate the viability of implementing transfer learning for cross-subject recognition.

LSTM models pre-trained as FINs outperforms other models by at least 1.06\% - 2.16\%. In constrast, LSTM+CNN-II (which excludes FIN pretraining) underperforms by $>$14.3\% compared to models using ground-truth features. This suggests that imitating established features reduces the search space \cite{phinyomark2018emg}, making it easier for the model to optimize representations for the downstream classification task.

\subsubsection{Model Training Time}

CNN-I (multi-subject) is significantly slower to converge than CNN-II (single-subject) ($31$ epochs $\times12s$ vs. $12$ epochs $\times0.46s$), suggesting that temporal features vary considerably across subjects (Table \ref{tab:model_speed}). The LSTM+CNN architecture requires multiple training repetitions, while our LSTM-FIN+CNN-II approach only needs fine-tuning repetitions, potentially saving computational resources. LSTM-FIN pre-training converges rapidly in 3 epochs (98 seconds each), demonstrating the efficiency of our feature imitation approach. These findings highlight the advantages of our proposed method in terms of learning efficiency and computational resource utilization.

\begin{table}[b]
    \caption{Comparison of Model Convergences}
    \label{tab:model_speed}
    \centering
    \small
    \setlength{\tabcolsep}{12pt}
    \begin{tabular}{|l|c|c|}
        \hline
        \textbf{Model} & \textbf{\# Epochs} & \textbf{Epoch time}\\
        \hline
        CNN-I &  31 & 12s\\
        CNN-II & 12 & 0.46s\\
        LSTM+CNN & 5 & 291s\\
        LSTM-FIN+CNN-II  & 4 & 58s\\
        LSTM-FIN & 3 & 98s \\
        \hline
    \end{tabular}
\end{table}

\subsection{Exp. \#4: Data Efficiency Validation}

Varying the proportion of fine-tuning data (from 20\% to 100\%) yielded a flat trend (Fig. \ref{fig:fin_cnn_percentage} - top), indicating that increasing the data from existing repetitions has minimal impact on the accuracy. It is important to note that the across-subjects set used nearly 3 times more data than the within-subjects set for the same percentage due to additional repetitions (additional repetitions are 1, 3, 4). However, even though across-subjects contained more repetitions, 100\% data usage of the within-subject set is still more than 20\% of data usage of the across-subject set. In summary, incorporating more repetitions is far more effective than stacking data from single repetitions; on the other hand, the accuracy is less subjective to the data usage proportion, which guarantees the performance under scarce labeled data circumstances.

\begin{figure}[h!]
\centering
    \begin{minipage}[b]{0.5\textwidth}
    \centering
    \includegraphics[width=1\textwidth]{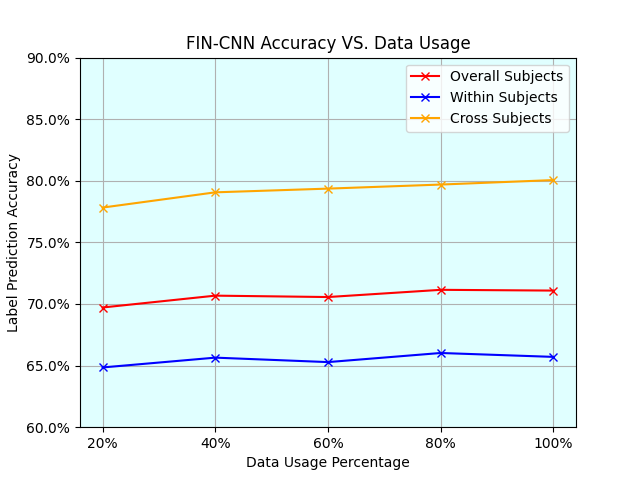}
    \end{minipage}
    \hfill
    \begin{minipage}[b]{0.5\textwidth}
    \centering
    \includegraphics[width=0.95\textwidth]{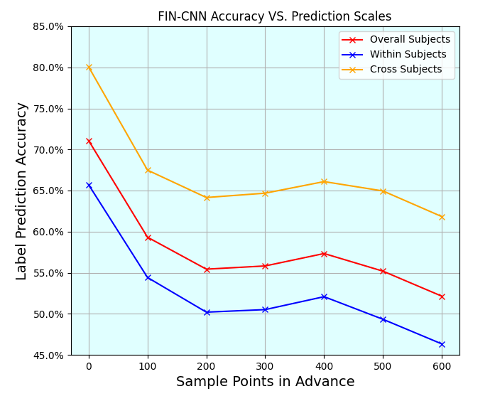}
    \end{minipage}
\caption{LSTM-FIN+CNN-II Model Performance Across Data Sizes and Future Predictions. (Top) Illustrates the mean accuracy given various proportions of training data. (Bottom) Illustrates the mean accuracy given variations in time to future event (100 samples = 50ms).}
\label{fig:fin_cnn_percentage}
\end{figure}



\subsection{Exp. \#5: Forward-in-time Classification}
In our experiments, we also considered the possibility of predicting future actions (Fig. \label{fig:fin_cnn_percentage} - bottom). We explored the capability of LSTM-FIN+CNN-II to predict future actions by varying the prediction window stride from 50ms to 300ms, in 50ms increments. The model demonstrated a roughly 15\% accuracy drop in low latency scenarios, indicating decent performance in near-future movement prediction. 

\section{Conclusion}
\label{sec:conclusion}
In this paper, we explored the application of Feature Imitation Networks (FINs) to represent four temporal sEMG signal features. Our LSTM-based FIN accurately simulated these features with R² $>$ 0.96 and MAP $>$ 88\%. The CNN classifier with feature augmentation outperformed traditional SVM classifiers, and when combined with the LSTM-FIN, it surpassed both standalone CNN (trained directly on features) and LSTM-CNN architectures (trained from the raw sEMG). Our experiments with varying data sizes demonstrated the model's robustness under limited labeled data conditions. We also found that incorporating more repetitions was crucial for improving accuracy. The model also showed potential for predicting movements up to 300ms in the future, albeit with lower accuracy compared to real-time classification.

This work highlights the effectiveness of pre-trained FIN models in transfer learning scenarios, addressing challenges related to data scarcity and simulated low-latency environments. The LSTM-FIN+CNN approach balances generalization and subject-specific adaptation, showing promise for practical sEMG-based movement recognition applications.

For future work, we propose exploring (1) the use of other neural network structures (e.g. CNN, feedforward) as FINs, (2) learning to imitate other time-frequency features, with the aim of creating robust and interpretable neural network models. We propose (3) investigating transfer learning between intact and amputated populations, a challenging transfer learning task, as well as (4) evaluating the forward-in-time classification model with end-users to assess the responsiveness of our approach to real-world cases.

\small
\bibliographystyle{IEEEbib}
\bibliography{refs}



\end{document}